\documentstyle[12pt]{article}
\newcommand{\bea}{\begin{eqnarray}}
\newcommand{\eea}{\end{eqnarray}}
\newcommand{\be}{\begin{equation}}
\newcommand{\ee}{\end{equation}}
\newcommand{\p}{\prime}
\newcommand{\nn}{\nonumber}
\newcommand{\rf}[1]{(\ref{#1})}

\begin{document}

\begin{center}

{\Large\bf SECOND-CLASS CONSTRAINTS AND LOCAL SYMMETRIES}\\
\vspace*{1cm}
{\large N.P.Chitaia, S.A.Gogilidze}\\
{\it Tbilisi State University, Tbilisi, University St.9,
380086 Georgia,} \\
\bigskip
and \\
\bigskip
{\large{Yu.S.Surovtsev}} \\
{\it Bogoliubov Laboratory of Theoretical Physics, Joint Institute for Nuclear
Research,
Dubna 141 980, Moscow Region, Russia}
\end{center}
\begin{abstract}
In the framework of the generalized Hamiltonian formalism by Dirac, the local
symmetries of dynamical systems with first- and second-class constraints are
investigated. For theories with an algebra of constraints of special form (to
which a majority of the physically interesting theories belongs) the method of
constructing the generator of local-symmetry transformations is obtained from
the requirement of the quasi-invariance of an action. It is proved that
second-class constraints do not contribute to the transformation law of the
local symmetry which entirely is stipulated by all the first-class
constraints. It is thereby shown that degeneracy of special form theories with
the first- and second-class constraints is due to their quasi-invariance under
local-symmetry transformations.
\end{abstract}
\section{Introduction}
In his basic works \cite{Dirac} on the generalized Hamiltonian formalism Dirac
has shown that from the presence of first-class constraints in a theory the
existence of the local symmetry group follows, a rank of which is determined
by the number of first-class primary constraints. In the same place it is
pointed out that,``possibly, all the first-class secondary constraints are to
be attributed to a class of generators of transformations which are not
related with a change of the physical state'' (Dirac's hypothesis). In
connection with the importance of constructing gauge transformations this
hypothesis has brought about a rather excited discussion
\cite{Shanmu}-\cite{GSST:JP}. In papers \cite{Cawley}-\cite{Stefano} it is
queried. And in refs.\cite{Sugano-Kimura}-\cite{Lusanna} one even asserts that
second-class constraints contribute also to a generator of gauge
transformations which become global in the absence of first-class constraints
\cite{Sugano-Kimura}. The generalized Hamiltonian dynamics of systems with
constraints of first and second class is at all studied relatively weakly up
to now. For example, only recently there have appeared the real schemes of
separation of constraints into the first- and second-class ones
\cite{Batlle-GPR}-\cite{CGS-1}. Explicit form of the local-symmetry
transformations is needed in both the traditional Dirac approach and, e.g.,
for realization of the presently popular BRST-BFV methods of covariant
quantization \cite{Frad-Vilk}-\cite{Henneaunex}.

In our previous papers \cite{GSST}-\cite{GSST:JP}, we have suggested a
method of constructing the generator of gauge transformations for singular
Lagrangians only with first-class constraints. At present, we extend our
scheme also to the theories with second-class constraints; and moreover, in
the given work we consider theories with an algebra of constraints of special
form, when first-class primary constraints are the ideal of quasi-algebra of
all the first-class constraints. A majority of the physically interesting
theories satisfy this condition. The general case (without restrictions on
the algebra of constraints) will be investigated in a subsequent paper.
The local-symmetry transformations is looked for here from the requirement of
a quasi-invariance (within a surface term) of the action functional under
these transformations. To elucidate a role of second-class constraints in the
local-symmetry transformations, we consider first- and second-class
constraints on the same basis in the hypothetical generator of these
transformations. We prove that the second-class constraints do not contribute
to the local-symmetry transformation law and, thus, the transformation
generator is a linear combination only of all the first-class constraints.

The paper is organized as follows. Section 2 is devoted to constructing the
local-symmetry transformation generator in the theories with constraints of
first and second class and to proving of that the latter are in no way
responsible for this symmetry. These derivations are based substantially on
results of our previous paper \cite{CGS-1} (below cited as paper I) on the
separation of constraints into the first- and second-class ones and on
properties of the canonical set of constraints. In the 3rd section our results
are examplified with a number of model Lagrangians \cite{Sugano-Kimura}, the
Chern--Simons theory and spinor electrodynamics.

\section{Local Symmetry Transformations}

Let us consider a dynamical system with the canonical set $(\Phi_\alpha^
{m_\alpha},\Psi_{a_i}^{m_{a_i}})$ of first- and second-class constraints,
respectively $(\alpha=1,\cdots,F,~m_\alpha=1,\cdots,M_\alpha;~~a_i=1,\cdots,
A_i,~m_{a_i}=1,\cdots,M_{a_i},~i=1,\cdots,n)$. Passing to this set from the
initial one is always possible in an arbitrary case by the method developed
in paper I.

A group of phase-space coordinate transformations, that maps each solution
of the Hamiltonian equations of motion into the solution of the same
equations, will be called the symmetry transformation. Under these
transformations the action functional is quasi-invariant within a surface
term.

Consider the action
\be \label{S}
S=\int_{t_1}^{t_2}dt~(p\dot q - H_T),
\ee
where
\be \label{H_T}
H_T = H + u_\alpha\Phi_\alpha^1,
\ee
$$H = H_c+\sum_{i=1}^{n}({\bf K}^{1~i})_{b_i~a_i}^{-1}\{\Psi_{a_i}^i,H_c\}
\Psi_{b_i}^1$$
is a first-class function \cite{Dirac}, $H_c$ is the canonical Hamiltonian,
$u_\alpha$ are the Lagrange multipliers.

We shall require a quasi-invariance of the action $S$ with respect to
transformations:
\bea \label{q-q^prime}
\left\{\begin{array}{ll} q_i^\p = q_i+\delta q_i ,& \quad
\delta q_i = \{q_i , G\} ,
\\ p_i^\p = p_i+\delta p_i ,& \quad\delta p_i =
\{p_i , G\} .  \end{array}\right.
\eea
The generator $G$ will be looked for in the form
\be \label{G-Dirac}
G=\varepsilon_\alpha^{m_\alpha}\Phi_\alpha^{m_\alpha} +
\eta_{a_i}^{m_{a_i}}\Psi_{a_i}^{m_{a_i}}.
\ee
In contrast to our previous works, in expression \rf{G-Dirac} for $G$ the
second term with constraints $\Psi_{a_i}^{m_{a_i}}$ is added, because we will
elucidate a role of second-class constraints under these transformations.

So, under transformations \rf{q-q^prime} we have
\bea \label{delta-S}
\delta S=\int_{t_1}^{t_2}dt\bigl[\delta p~\dot q+p~\delta\dot q-
\delta H_T\bigr]&=&\int_{t_1}^{t_2}dt~\bigl[\frac{d}{dt}(p~\frac{\partial G}
{\partial p}-G) \nn\\
&&~~~~~~~~~+\frac{\partial G}{\partial t}+\{G,H_T\}\bigr].
\eea
>From \rf{delta-S} we see: in order that the transformations \rf{q-q^prime}
were the symmetry ones, it is necessary
\be \label{sym-condition}
\frac{\partial G}{\partial t}+\{G,H_T\}~\stackrel{\Sigma_1}{=}0.
\ee
That the last equality must be realized on the primary-constraint surface
$\Sigma_1$, can be easily interpreted if one remembers that the surface
$\Sigma_1$ is the whole $(q,\dot q)$-space image in the phase space. Since
under the operation of the local-symmetry transformation group the
$(q,\dot q)$-space is being mapped into itself in a one-to-one manner,
therefore the one-to-one mapping of $\Sigma_1$ into itself corresponds to this
in the phase space. Therefore, at looking for the generator $G$ it is
natural to require also the primary-constraint surface $\Sigma_1$ to be
conserved under transformations \rf{q-q^prime} and \rf{G-Dirac}, i.e. the
requirement \rf{sym-condition} must be supplemented by the demands
\bea
&& \bigl\{G,\Psi_{a_i}^1\bigr\}~\stackrel{\Sigma_1}{=}0, \label{PB-G-Psi1} \\
&& \bigl\{G,\Phi_\alpha^1\bigr\}~\stackrel{\Sigma_1}{=}0, \label{PB-G-Phi1}
\eea
Note in connection with the relations \rf{PB-G-Psi1} and \rf{PB-G-Phi1} that
the symmetry group of the action functional for dynamical system is the
symmetry group of the motion equations obtained from the variational principle
(the inverse is incorrect in the general case). Since the constraint equations
$\Phi_\alpha^1=0$ and $\Psi_{a_i}^1=0$ are contained in a system of the motion
equations, the relations \rf{PB-G-Psi1} and \rf{PB-G-Phi1} are an expression
of this group property \footnote{The relations of type \rf{PB-G-Psi1} and
\rf{PB-G-Phi1} on the secondary constraints are not imposed in accordance with
reasoning after formula \rf{sym-condition}.}.

Furhter we shall use the following Poisson brackets among the canonical
constraint set $(\Phi,\Psi)$ and $H$ established in paper I:
\bea
&& \bigl\{\Phi_\alpha^{m_\alpha},H\bigr\} =
g_{\alpha~~\beta}^{m_\alpha m_\beta}~\Phi_\beta^{m_\beta},\quad
m_\beta=1,\cdots,m_\alpha+1,~~~~~~~~~~~~~~~~~~~ \label{PB-Phi-H^prime}\\
&& \bigl\{\Psi_{a_i}^{m_{a_i}},H\bigr\} = \bar{g}_{{a_i}~~\alpha}^{m_{a_i}
m_\alpha}~\Phi_\alpha^{m_\alpha}+\sum_{k=1}^{n}h_{{a_i}~~{b_k}}^{m_{a_i}
m_{b_k}}~\Psi_{b_k}^{m_{b_k}},\quad m_{b_n}=m_{a_i}+1,~~~~~~~\label{PB-Psi-H^prime}\\
&& \bigl\{\Phi_\alpha^{m_\alpha},\Phi_\beta^{m_\beta}\bigr\} =
f_{\alpha~~\beta~~\gamma}^{m_\alpha m_\beta m_\gamma}~\Phi_\gamma^{m_\gamma},
~~~~~~~~~~~~~~~~~~~\label{PB-Phi-Phi}\\
&& \bigl\{\Psi_{a_i}^{m_{a_i}},\Psi_{b_k}^{m_{b_k}}\bigr\} =\bar{f}_
{{a_i}~~{b_k}~~\gamma}^{m_{a_i} m_{b_k} m_\gamma}~\Phi_\gamma^{m_\gamma}+
\sum_{l=1}^{n}k_{{a_i}~~{b_k}~~{c_l}}^{m_{a_i} m_{b_k} m_{c_l}}~\Psi_{c_l}^
{m_{c_l}}+D_{{a_i}~~{b_k}}^{m_{a_i} m_{b_k}},~~~~~~~ \label{PB-Psi-Psi}
\eea
where the structure functions, generally speaking, depend on $q$ and $p$ and,
besides, one can see that
\be \label{g}
g_{\alpha~~\beta}^{m_\alpha m_\beta}=0,\quad\mbox{if}~m_\alpha+2\leq m_\beta~,
\ee
\bea \label{bar-g-h}
\left\{\begin{array}{ll}
\bar{g}_{{a_i}~~\alpha}^{m_{a_i} m_\alpha}=0,&~~\mbox{if}~~ m_\alpha\geq
m_{a_i}~,\\{}\\ h_{{a_i}~~{b_k}}^{m_{a_i} m_{b_k}}=0,&~~\mbox{if}~~ m_{a_i}+2
\leq m_{b_k}~~\mbox{or if}~~a_i=b_k,~~m_{a_i}=M_{a_i},\\
{} &~~m_{b_k}\geq M_{a_i}, \end{array}\right.
\eea
\be \label{f}
f_{\alpha~~\beta~~\gamma}^{1~ m_\beta m_\gamma}=0~~~\mbox{for}~~
m_\gamma\geq 2,
\ee
\bea \label{F-sym}
\left\{\begin{array}{l}
F_{{a_i}~~~~~{b_i}}^{M_{a_i}-l~l+1}=(-1)^l~F_{{a_i}~{b_i}}^{1~M_{b_i}}~,~~
~~l=0,1,\cdots,M_{a_i}-1,\\ {}\\
F_{{a_i}~{b_i}}^{j~k}=0,~~~~~~~\mbox{if}~~j+k\neq M_{a_i}+1,\\ {}\\
F_{{a_i}~~{b_k}}^{m_{a_i} m_{b_k}}=0,~~~~\mbox{if}~~a_i,b_k~\mbox
{refer to different chains (or doubled}\\ ~~~~~~~~~~~~\mbox{chains) of
second-class constraints}~
\bigl(D_{{a_i}~~{b_k}}^{m_{a_i} m_{b_k}}~\stackrel{\Sigma}{=}F_{{a_i}~~{b_k}}^
{m_{a_i} m_{b_k}}\bigr).
\end{array}\right.
\eea
The equality \rf{f} reflects the first-class primary constraints to make a
subalgebra of quasi-algebra of all the first-class constraints. The equalities
\rf{F-sym} express partly the structure of the canonical second-class
constraints established in paper I.

So, from eqs.\rf{sym-condition} and \rf{G-Dirac} with taking account of
\rf{PB-Phi-H^prime}-\rf{PB-Psi-Psi} we write down
\bea \label{eps.eta-Phi-Psi}
&&\Bigl (\dot \varepsilon_\alpha^{m_\alpha}+\varepsilon_\beta^{m_\beta}
g_{\beta~~\alpha}^{m_\beta m_\alpha}+\sum_{i=1}^{n}\eta_{a_i}^{m_{a_i}}
\bar{g}_{{a_i}~~\alpha}^{m_{a_i} m_\alpha}~\Bigr)\Phi_\alpha^{m_\alpha}\nn\\
&& +\sum_{i=1}^{n}\Bigl(\dot{\eta}_{a_i}^{m_{a_i}}+\sum_{k=1}^{n}\eta_{b_k}^
{m_{b_k}} h_{{b_k}~~{a_i}}^{m_{b_k} m_{a_i}}~\Bigr)\Psi_{a_i}^{m_{a_i}} \\
&& +u_\alpha\bigl\{G,\Phi_\alpha^1\bigr\} ~\stackrel{\Sigma_1}{=}0.  \nn
\eea
Taking into consideration \rf{PB-G-Phi1}, we have
\be \label{cond.PB-G-Phi1}
u_\alpha\bigl\{G,\Phi_\alpha^1\bigr\} ~\stackrel{\Sigma_1}{=}0.
\ee
Then, in view of the functional independence of constraints $\Phi_\alpha^
{m_\alpha}$ and $\Psi_{a_i}^{m_{a_i}}$, in order to satisfy the equality
\rf{eps.eta-Phi-Psi} one must demand the coefficients of constraints
$\Phi_\alpha^{m_\alpha}~(m_\alpha \geq 2)$ and $\Psi_{a_i}^{m_{a_i}}~(m_{a_i}
\geq 2)$ to vanish. Note that, even if not all the constraints are
functionally independent, the vanishing of the coefficients of constraints
$\Phi_\alpha^{m_\alpha}$ in \rf{eps.eta-Phi-Psi} ensures the quasi-invariance
of the action functional at the assumption that second-class constraints do
not contribute to the transformations \rf{q-q^prime}. However, to investigate
the role of second-class constraints, it is convenient to consider the
functional independence of constraints, since otherwise one can always pass to
an equivalent set of functionally-independent constraints, for example, by the
proper Abelianization procedure \cite{GKP}.

So, before analyzing these conditions to satisfy the equality
\rf{eps.eta-Phi-Psi}, let us consider in detail the conditions of the
primary-constraints surface conservation starting from \rf{PB-G-Psi1}. Its
realization would mean the presence of the following equalities:  \be
\label{PB-G-Psi1-cond}
\varepsilon_\alpha^{m_\alpha}\bigl\{\Phi_\alpha^{m_\alpha},\Psi_{a_i}^1\bigr\}
~\stackrel{\Sigma_1}{=}0, \qquad
\sum_{k=1}^{n}\eta_{b_k}^{m_{b_k}}\bigl\{\Psi_{b_k}^{m_{b_k}},\Psi_{a_i}^1
\bigr\}~\stackrel{\Sigma_1}{=}0.
\ee
The first requirement \rf{PB-G-Psi1-cond} may be always realized by vanishing
the Poisson brackets with the help of the corresponding transformation of
equivalence as it is made in our previous paper I.

Since we take that passing to the canonical constraints set of paper I has
been performed, in the second equality \rf{PB-G-Psi1-cond} for each value of
$a_i$ in the double sum over $k$ and over $b_k$ the only non-vanishing Poisson
brackets are those at $b_k=a_i,~M_{b_k}=i$, therefore
\be \label{eta-i=0}
\eta_{a_i}^i=0\quad \mbox{for}~ i=1,\cdots,n,
\ee
i.e. we have determined that in expression \rf{G-Dirac} the coefficients of
those $i$-ary constraints, which are the final stage of each chain of second
class constraints, and of those second-class primary constraints, which do not
generate the secondary constraints, disappear. Now we consider the requirement
of vanishing the coefficients of constraints $\Psi_{a_i}^{m_{a_i}}~(m_{a_i}
\geq 2,~i=2,\cdots,n,~a_i=1,\cdots,A_i)$ in eq. \rf{eps.eta-Phi-Psi}:
\bea \label{eq:eta}
\left\{\begin{array}{l}
\dot{\eta}_{a_n}^n + \eta_{b_n}^n h_{b_n~a_n}^{n~~n} + \eta_{b_n}^{n-1}
h_{b_n~~a_n}^{n-1~n} =0,\\
{}\\
\dot{\eta}_{a_n}^{n-1} + \eta_{b_n}^n h_{b_n~a_n}^{n~~n-1} + \eta_{b_n}^{n-1}
h_{b_n~~a_n}^{n-1~n-1} + \eta_{b_n}^{n-2}h_{b_n~~a_n}^{n-2~n-1}=0,\\
~\vdots\\
\dot{\eta}_{a_2}^2 + \eta_{b_2}^2 h_{b_2~a_2}^{2~~2} + \eta_{b_2}^1
h_{b_2~a_2}^{1~~2} =0.
\end{array}\right.
\eea
In this system of equations the number of unknown functions exceeds the number
of equations by the number of the second-class primary constraints which make
up the constraint chains. However, we have already established the result
\rf{eta-i=0} for senior terms of each subsystem of equations for $a_i$.
Inserting the values $\eta_{a_n}^n=0$ into the first line of system
\rf{eq:eta}, we obtain a system of $A_n$ algebraic linear homogeneous
equations for $A_n$ unknowns $\eta_{b_n}^{n-1}$ that has only a trivial
solution $\eta_{b_n}^{n-1}=0$ since $\mbox{det}\|h_{b_n~~a_n}^{n-1~n}\|\neq
0$~ (see below). Using this result in the second line of \rf{eq:eta}, we
obtain a system of analogous equations for unknowns $\eta_{b_n}^{n-2}$. Its
solution is $\eta_{b_n}^{n-2}=0$ since
$\mbox{det}\|h_{b_n~~a_n}^{n-2~n-1}\|\neq 0$. Continuing successively this
process we shall deduce that all quantities $\eta_{a_i}^{m_{a_i}}$ vanish,
i.e. the second-class constraints do not contribute to the generator of
local-symmetry transformations.

Now we shall show that
\be \label{det-not-0}
\mbox{det}\|h_{b_i~~~~~a_i}^{i-k-1~i-k}\|\neq 0
\ee
since a set of all constraints consists of independent functions. We shall
apply a method by contradiction, i.e. suppose the indicated determinant to
vanish. Further from the relation \rf{PB-Psi-H^prime} we have
\be \label{PB:Psi-{i-k-1}-H}
\bigl\{\Psi_{a_i}^{i-k-1},H_c\bigr\}~\stackrel{\Sigma_{i-k-1}}{=}~
h_{a_i~~~~~b_m}^{i-k-1~i-k}~\Psi_{b_m}^{i-k}
\ee
where $\Sigma_{i-k-1}$ is the surface of all constraints up to and including
the ${i-k-1}$ stage. But the assumption of vanishing the determinant of
the matrix $\|h_{b_i~~~~~a_i}^{i-k-1~i-k}\|$ means a linear dependence of its
some rows or columns:
\be \label{lin-depend-of-column}
h_{b_i~~~~~a_m}^{i-k-1~i-k}=C_{a_is_i}h_{s_i~~~~~b_m}^{i-k-1~i-k}.
\ee
Inserting \rf{lin-depend-of-column} into the right-hand side of
\rf{PB:Psi-{i-k-1}-H} we obtain
$$\bigl\{\Psi_{a_i}^{i-k-1},H_c\bigr\}~\stackrel{\Sigma_{i-k-1}}{=}~
C_{a_is_i}h_{s_i~~~~~b_m}^{i-k-1~i-k}~\Psi_{b_m}^{i-k}~\stackrel{\Sigma_
{i-k-1}}{=}~\bigl\{C_{a_is_i}\Psi_{s_i}^{i-k-1}~,H_c\bigr\}.$$
>From here write down
$$\bigl\{\Psi_{a_i}^{i-k-1}-C_{a_is_i}\Psi_{s_i}^{i-k-1}~,H_c\bigr\}
~\stackrel{\Sigma_{i-k-1}}{=}~0.$$
The last equality means that
$$\Psi_{a_i}^{i-k-1}=C_{a_is_i}\Psi_{s_i}^{i-k-1}+c_{a_i}^{m_{a_i}}\Psi_{a_i}
^{i-m_{a_i}} + d_\alpha^{m_\alpha}\Phi_\alpha^{i-m_\alpha},$$
$$m_{a_i}=k+1,\cdots,i-1,~~~~m_\alpha=k+1,\cdots,i-1,$$
where $c_{a_i}^{m_{a_i}}$ and $d_\alpha^{m_\alpha}$ are arbitrary functions of
$q$ and $p$, non-vanishing on the constraint surface $\Sigma_{i-k-1}$. Thus,
we have arrived at the contradiction with the condition of independence of
constraints. This proves the validity of \rf{det-not-0}.

Returning to the second condition \rf{PB-G-Phi1} of the primary-constraint
surface conservation under local-symmetry transformations, we see that it (and
from here, too, the equality \rf{cond.PB-G-Phi1}) will be fulfilled if
\be \label{ideal}
\bigl\{\Phi_\alpha^1,\Phi_\beta^{m_\beta}\bigr\}=f_{\alpha~~\beta~~\gamma}^
{1~~ m_\beta~ 1}~\Phi_\gamma^1 .
\ee
This relation emerged already earlier \cite{GSST} in the case of dynamical
systems only with the constraints of first class and ensured the conservation
of the primary-constraint surface $\Sigma_1$ under the local-symmetry
transformations. Here it means a quasi-algebra of special form where the
first-class primary constraints make an ideal of quasi-algebra formed by all
first-class constraints, also in the presence of second-class constraints.

To determine the multipliers $\varepsilon_\alpha^{m_\alpha}$ in the generator
\rf{G-Dirac}, now we have only the requirement of vanishing the coefficients
of constraints $\Phi_\alpha^{m_\alpha}$ in \rf{eps.eta-Phi-Psi} \cite{GSST}:
\be \label{eq:eps}
\dot \varepsilon_\alpha^{m_\alpha}+\varepsilon_\beta^{m_\beta}
g_{\beta~~\alpha}^{m_\beta m_\alpha}=0,\qquad m_\beta= m_\alpha-1,\cdots,
M_\alpha .
\ee
In the system of equations \rf{eq:eps}, the number of unknowns exceeds the
number of equations by the number $F=A-\sum_{i=1}^{n}A_i$ of the first-class
primary constraints, therefore the system \rf{eq:eps} may be solved to within
$F$ arbitrary functions. This proves that the rank of the local-symmetry
transformation quasigroup is defined by the number of first-class primary
constraints also in the presence of second-class constraints. We shall remind,
for completeness, how one make use of this system of equations
\cite{GSST:tmf1}.  We write down \rf{eq:eps} as
\bea \label{eq:eps-detail}
&&\dot \varepsilon_\alpha^{M_\alpha} +\varepsilon_\beta^{M_\alpha}~g_{\beta~~~\alpha}^{M_\alpha~M_\alpha} +
\varepsilon_\beta^{M_\alpha-1}~g_{\beta~~~~~\alpha}^{M_\alpha-1~M_\alpha} = 0 ,~~~~~~~~~~~~~~~~~~~~~\nn\\
&&\dot \varepsilon_\alpha^{M_\alpha-1} + \varepsilon_\beta^{M_\alpha}~g_{\beta~~~\alpha}^{M_\alpha~M_\alpha-1} +
\varepsilon_\beta^{M_\alpha-1}~g_{\beta~~~~~\alpha}^{M_\alpha-1~M_\alpha-1}+
\varepsilon_\beta^{M_\alpha-2}~g_{\beta~~~~~\alpha}^{M_\alpha-2~M_\alpha-1} = 0 ,~~~\nn\\
&&~\vdots\\
&&\dot \varepsilon_\alpha^2 + \varepsilon_\beta^{M_\alpha}~g_{\beta~~~\alpha}^{M_\alpha~2} +
\varepsilon_\beta^{M_\alpha-1}~g_{\beta~~~~~\alpha}^{M_\alpha-1~2} + \cdots +
\varepsilon_\beta^1~g_{\beta~\alpha}^{1~2} = 0 ,~~~~~~~~~~~~~~~~~~~~~~\nn
\eea
$$\alpha,\beta = 1,\cdots, F .$$
Taking $\varepsilon_\alpha\equiv\varepsilon_\alpha^{M_\alpha}$ as arbitrary
functions and inserting them into the first line of system \rf{eq:eps-detail}
we obtain a system of $F$ inhomogeneous algebraic linear equations for $F$
unknowns $\varepsilon_\beta^{M_\alpha-1} (\beta = 1,\cdots, F)$. Solving this
system of equations (we have $\mbox{det}\|g_{\beta~~~~~\alpha}^{M_\alpha-1~M_
\alpha}\|\neq 0$ \cite{GSST:tmf1}) and inserting this result into the second
line of \rf{eq:eps-detail}, we obtain again a system of $F$ inhomogeneous
algebraic linear equations for $F$ unknowns $\varepsilon_\beta^{M_\alpha-2}$
that must be solved ($\mbox{det}\|g_{\beta~~~~~\alpha}^{M_\alpha-2~M_\alpha-1}
\|\neq 0$). The result must be inserted into the following line of system
\rf{eq:eps-detail}, etc., up to the last line which gives a system of $F$
equations for $F$ unknowns $\varepsilon_\beta^1$. Solving this last system of
equations we shall express, finally, all $\varepsilon_\alpha^{m_\alpha}$ in
terms of $\varepsilon_\alpha(t)$,$g_{\beta~~\alpha}^{m_\beta m_\alpha}$ and
their derivatives:
\be \label{epsilon}
\varepsilon_\alpha^{m_\alpha}=B_{\alpha~~\beta}^{m_\alpha m_\beta}
\varepsilon_\beta^{(M_\alpha-m_\beta)},\qquad m_\beta=m_\alpha ,\cdots,M_\alpha
\ee
(in formula \rf{epsilon} the summation runs also over $m_\beta$), where
$$\varepsilon_\beta^{(M_\alpha-m_\beta)}\equiv {{d^{M_\alpha-m_\beta}}\over{dt
^{M_\alpha-m_\beta}}}\varepsilon_\beta (t),\qquad \varepsilon_\beta (t)\equiv
\varepsilon_\beta^{M_\beta} $$
and $B_{\alpha~~\beta}^{m_\alpha m_\beta}$ are, generally speaking, functions
of $q$ and $p$ and their derivatives up to the order $M_\alpha-m_\alpha-1$.
Note that the condition
$$\mbox{det}\|g_{\beta~~~~~~~\alpha}^{M_\alpha-k-1~M_\alpha-k}\|\neq 0~~~
(k=0,1,\cdots,M_\alpha-2),$$
which is needed for the system of equations \rf{eq:eps} to have a solution, is
proved as a consequence of the functional independence of all constraints --
in the same way as in the case of dynamical systems with the constraints of
first class only \cite{GSST}, and in the same way as the similar condition
\rf{det-not-0} for the system of equations \rf{eq:eta}.
So, the generator of the local-symmetry transformations takes the form
\be \label{G}
G=B_{\alpha~~\beta}^{m_\alpha m_\beta}\phi_\alpha^{m_\alpha}\varepsilon_\beta^
{(M_\alpha-m_\beta)} , \qquad m_\beta=m_\alpha,\cdots,M_\alpha .
\ee
The obtained generator \rf{G} satisfies the group property
\be \label{G:group}
\{G_1,G_2\}=G_3,
\ee
where the transformation $G_3$ \rf{G} is realized by carrying out two
successive transformations $G_1$ and $G_2$ \rf{G}. The amount of group
parameters $\varepsilon_\alpha (t)$ which determine the rank of the quasigroup
of these transformations equals the number of primary constraints of first
class. As can be seen from formula \rf{G}, the transformation law may include
both arbitrary functions $\varepsilon_\alpha (t)$ and their derivatives up to
and including the order $M_\alpha-1$; the highest derivatives
$\varepsilon_\alpha^{(M_\alpha-1)}$ should be always present.

Thus, we have derived the generator of the local-symmetry transformations and
proved that there is no influence of second-class constraints on these
transformations from requirements of the action quasi-invariance and of
conservation of the primary-constraint surface under local-symmetry
transformations and on the basis of properties of the completely-separated
(into first- and second-class) constraint set.

Notice that the corresponding transformations of local symmetry in the
Lagrangian formalism are determined by following way:
\be \label{q-dot-q}
\delta q_i(t) =\{q_i(t), G\}\biggr|_{p=\frac{\partial L}{\partial
\dot q}},\qquad \delta\dot q(t) = \frac{d}{dt}\delta q(t).
\ee

\section{Examples}
In this section we illustrate our results by a number
of examples in both finite- and infinite-dimensional cases.

1. Consider the Lagrangian \cite{Sugano-Kimura}
\be \label{exam:L}
L=(\dot q_1 + \dot q_2) q_3+\frac{1}{2}{\dot q_3}^2-\frac{1}{2}{q_2}^2.
\ee
The generalized momenta are of the form:~$p_1=q_3,~p_2=q_3,~p_3=\dot q_3.$
Therefore we have two primary  constraints:
\be \label{exam:phi-prim}
\phi_1^1=p_1-q_3, \qquad \phi_2^1=p_2-q_3.
\ee
The total Hamiltonian gets the form:
\be \label{exam:H_T}
H_T=\frac{1}{2}({p_3}^2 + {q_2}^2) +u_1\phi_1^1+u_2\phi_2^1.
\ee
The self-consistency conditions of theory give
$$\dot \phi_1^1=\{\phi_1^1,H_T\}=-p_3,\qquad
\dot \phi_2^1=\{\phi_2^1,H_T\}=-q_2-p_3,$$
i.e. two secondary constraints
\be \label{exam:phi-sec}
\phi_1^2=p_3, \qquad \phi_2^2=p_3-q_2,
\ee
and
$$\dot \phi_1^2=\{\phi_1^2,H_T\}=u_1+u_2=0,\qquad
\dot \phi_2^1=\{\phi_2^1,H_T\}=u_2+u_1+u_2=0,$$
that means ~~~~$u_1=u_2=0.$
Two last equations serve for determining the Lagrangian multipliers $u_1$ and
$u_2$, and there no longer arise constraints. Let us calculate the matrix~
${\bf W}=\left\|{\bf K}^{m_\alpha m_\beta}\right\|=\left\|\{\phi_\alpha^
{m_\alpha},\phi_\beta^{m_\beta}\}\right\|$:
\be \label{exam:W}
{\bf W}=\left(\begin{array}{rrrr}
0 & ~~0 & -1 & -1 \\
0 & ~~0 & -1 & -2 \\
1 & ~~1 & 0 & 0 \\
1 & ~~2 & 0 & 0
\end{array}  \right).
\ee
We see that ~$\mbox{rank}{\bf W}=4$, i.e. all constraints are of second class,
therefore  ${\bf W}$  have a quasidiagonal (antisymmetric) form. Performing
our procedure we shall pass to the equivalent canonical set of constraints
$\Psi_a^ {m_a}$ according to the formula (57) of paper I:
\be \label{exam:Psi}
\left(\begin{array}{c} \Psi_1^1\\ \Psi_2^1\\ \Psi_1^2\\ \Psi_2^2 \end{array}
\right)= \left(\begin{array}{rrrr}
1 & 0 & ~~0 & 0 \\
1 & -1 & ~~0 & 0 \\
0 & 0 & ~~1 & 0 \\
0 & 0 & ~~1 & -1 \end{array} \right) \left(\begin{array}{c} \phi_1^1\\
\phi_2^1\\ \phi_1^2\\ \phi_2^2 \end{array}  \right)= \left(\begin{array}{c}
p_1-q_3\\ p_1-p_2\\ p_3\\ -q_2 \end{array}  \right).
\ee
For the last set of constraints the quasidiagonal form of  ${\bf W}$  will
have a canonical structure:
\be \label{exam:W^p} {\bf
W}^\p=\left(\begin{array}{rrrr}
0 & ~~0 & -1 & 0 \\
0 & ~~0 & 0 & -1 \\
1 & ~~0 & 0 & 0 \\
0 & ~~1 & 0 & 0
\end{array} \right).
\ee
Now for quasi-invariance of the action with respect to transformations
\rf{q-q^prime} with generator \rf{G-Dirac}
$$G=\eta_1^1~\Psi_1^1+\eta_2^1~\Psi_2^1+\eta_1^2~\Psi_1^2+\eta_2^2~\Psi_2^2,$$
it is necessary to realize the condition \rf{PB-G-Psi1} of conservation of the
primary-constraint surface under these transformations
\be \label{exam:PB-G-Psi1}
\bigl\{G,\Psi_a^1\bigr\}~\stackrel{\Sigma_1}{=}0,\qquad a=1,2.
\ee
>From \rf{exam:PB-G-Psi1} we obtain ~$\eta_1^2=\eta_2^2=0.$  Next from
\rf{eq:eta} we establish ~$\eta_1^1=\eta_2^1=0,$  i.e. the second-class
constraints of system \rf{exam:Psi} generate the transformations of neither
local symmetry nor global one.

2. Consider the Lagrangian \cite{Sugano-Kimura}
\be \label{exam2:L}
L=\dot q_1 q_2-\dot q_2 q_1-(q_1-q_2) q_3.
\ee
Then passing to the Hamiltonian formalism we obtain the generalized momenta
~$p_1=q_2, \quad p_2=-q_1, \quad p_3=0$~ and, thus, three primary constraints:
\be \label{exam2:phi-prim}
\phi_1^1=p_1-q_2, \qquad \phi_2^1=p_2+q_1, \qquad \phi_3^1=p_3.
\ee
The total Hamiltonian gets the form:
\be \label{exam2:H_T}
H_T=(q_1-q_2) q_3 +u_1\phi_1^1+u_2\phi_2^1+u_3\phi_3^1.
\ee
>From the self-consistency conditions of the theory we obtain
\be \label{exam2:consist.cond.}
\dot \phi_1^1=-q_3-2u_2=0,\quad \dot \phi_2^1=q_3+2u_1=0, \quad \dot
\phi_3^1=-q_1+q_2=0.
\ee
Two first equations \rf{exam2:consist.cond.} serve for determining the
Lagrangian multipliers: ~$u_1=u_2=-\frac{1}{2}q_3.$~ The last relation
\rf{exam2:consist.cond.} gives the secondary constraint $$\phi_3^2=q_2-q_1,$$
and there no longer arise constraints. Let us calculate the matrix~ ${\bf W}=
\left\|\{\phi_\alpha^ {m_\alpha},\phi_\beta^{m_\beta}\}\right\|$:
\be \label{exam2:W}
{\bf W}=\left(\begin{array}{rrrr}
0 & -2 & ~~0 & 1 \\
2 & 0 & ~~0 & -1 \\
0 & 0 & ~~0 & 0 \\
-1 & 1 & ~~0 & 0
\end{array}  \right).
\ee
>From ~$\mbox{rank}{\bf W}=2$~ we conclude that two constraints are of second
class and two ones are of first class. With the help of our procedure we
separate constraints into those of first and second class. For this purpose,
by means of the equivalence transformation we pass to the canonical set of
constraints according to the formula (57) of paper I:
\be \label{exam2:Psi-Phi}
\left(\begin{array}{c}
\Psi_1^1\\ \Psi_2^1\\ \Phi_1^1\\ \Phi_1^2 \end{array} \right)=
\left(\begin{array}{rrrr}
1 & 0 & 0 & 0 \\
0 & 1 & 0 & 0 \\
0 & 0 & 1 & 0 \\
1 & 1 & 0 & 2 \end{array} \right)
\left(\begin{array}{c}
\phi_1^1\\ \phi_2^1\\
\phi_3^1\\ \phi_3^2 \end{array} \right)= \left(\begin{array}{c}
p_1-q_2\\p_2+q_1\\ p_3\\ p_1+p_2+q_2-q_1 \end{array}  \right).
\ee
For the last set of constraints the matrix ${\bf W}$ acquires the canonical
form:
\be \label{exam2:W^p}
{\bf W}^\p=\left(\begin{array}{rrrr}
0 & -2 & ~~0 & ~~0 \\
2 & 0 & ~~0 & ~~0 \\
0 & 0 & ~~0 & ~~0 \\
0 & 0 & ~~0 & ~~0
\end{array} \right).
\ee
Further we look for the generator $G$ in the form \rf{G-Dirac}:
\be \label{exam2:G-Dirac}
G=\eta_1^1~\Psi_1^1+\eta_2^1~\Psi_2^1+\varepsilon_1^1~\Phi_1^1+
\varepsilon_1^2~\Phi_1^2.
\ee
>From the second condition \rf{PB-G-Psi1} of conservation of the
primary-constraint surface $\Sigma_1$ under transformations \rf{q-q^prime} we
derive ~$\eta_1^1=\eta_2^1=0,$  i.e. the second-class constraints of the
system do not contribute to the generator $G$. The first condition
\rf{PB-G-Phi1} of conservation of $\Sigma_1$ is realized because
$$\bigl\{\Phi_1^1,\Phi_1^2\bigr\}=0.$$
If we take into account that ~$g_{1~1}^{1~2}=\frac{1}{2}$~ and
~$g_{1~1}^{2~2}=0$~ in \rf{PB-Phi-H^prime}, equation \rf{eq:eps} becomes
$$\dot{\varepsilon}_1^2 + \frac{1}{2}\varepsilon_1^1 = 0.$$
Denoting ~$\varepsilon_1^2\equiv \varepsilon$,~ we obtain ~$\varepsilon_1^1=
-2\dot{\varepsilon}$~ and, therefore,
$$G=-2\dot{\varepsilon}p_3+\varepsilon(p_1+p_2+q_2-q_1),$$
which gives ~$\delta q_1=\varepsilon,~ \delta q_2=\varepsilon,~ \delta q_3=
-2\dot{\varepsilon},~ \delta p_1=\varepsilon,~ \delta p_2=-\varepsilon,~
\delta p_3=0.$\\
In the $(q,\dot q)$-space the local-symmetry transformations are established
with the help of formulas \rf{q-dot-q}. It is easy to verify that the action
is invariant with respect to the transformations generated by $G$. This is a
consequence of the constraints being linear in the momentum variables.

3. We now look at the infinite-dimensional cases. We consider first a
Chern--Simons theory. Theories of such type describe, e.g., the fractional
quantum Hall effect and other phenomena.

The Lagrangian density for a complex field $\phi$ interacting with an Abelian
Chern--Simons field is \cite{Semenoff}
\be \label{Chern:L}
{\cal L} = (\partial_\mu+iA_\mu) \varphi^* (\partial^\mu-iA^\mu) \varphi+\frac
{\alpha}{4\pi}\varepsilon_{ij} \Bigl(A_0 \partial_i A_j + \dot A_i A_j
+ A_i \partial_j A_0 \Bigr),
\ee
where ~$i,j=1,2$~ and ~$\mu=0,1,2.$~ The generalized momenta are
$$\pi_0=\frac{\partial {\cal L}}{\partial \dot A_0}=0, \qquad
\pi_i=\frac {\partial {\cal L}}{\partial \dot A_i}=\frac{\alpha}{4\pi}
\varepsilon_{ij} A_j, $$
$$\pi_\varphi=\frac{\partial {\cal L}}{\partial \dot \varphi}=(\partial_0+
iA_0)\varphi^*, \qquad \pi_{\varphi^*}=\frac{\partial {\cal L}}{\partial \dot
\varphi^*}=(\partial_0-iA_0) \varphi.$$
Therefore, in the phase space we have three primary constraints:
\be \label{Chern:phi-prim}
\phi_i^1=\pi_i-\frac{\alpha}{4\pi} \varepsilon_{ij}A_j,\quad i,j=1,2, \qquad
\phi_3^1=\pi_0
\ee
and the canonical Hamiltonian:
\bea \label{Chern:H_T}
H_c &=& \int d^2x
\Bigl[\pi_\varphi(x)\pi_{\varphi^*}(x)+\bigl(\partial_i+iA_i(x)\bigr)
\varphi^*(x)\bigl(\partial_i-iA_i(x)\bigr)\varphi(x)~~~~\nn\\
&&~~~~+A_0(x)j_\varphi-\frac{\alpha}{4\pi}\varepsilon_{ij} \Bigl(
A_0(x)\partial_i A_j(x)+A_i(x)\partial_j A_0(x)\Bigr)\Bigr],
\eea
where ~$j_\varphi=i\bigl(\varphi(x)\pi_\varphi(x)-\varphi^*(x)\pi_{\varphi^*}
(x)\bigr)$.

Among the conditions of the time conservation of constraints ~$\dot {\phi}_i^
1=0$\\ $(i=1,2)$~ and ~$\dot{\phi}_3^1=0$~ two first ones serve for determining
the Lagrangian multipliers $u_1$ and $u_2$:
$$u^1=\frac{4\pi}{\alpha}\bigl[i(\varphi\partial_2\varphi^*-\varphi^*
\partial_2\varphi)-2\varphi^*\varphi A_2\bigr]-2\partial_1 A_0,$$
$$u^2=\frac{4\pi}{\alpha}\bigl[i(\varphi^*\partial_1\varphi-\varphi\partial_1
\varphi^*)+2\varphi^*\varphi A_1\bigr]-2\partial_2 A_0.$$
>From the condition of conservation for $\phi_3^1$ we obtain the secondary
constraint
\be \label{Chern:phi_sec}
\phi_3^2= j_\varphi-\frac{\alpha}{2\pi}\varepsilon_{ij}\partial_i A_j,
\ee
and there do not arise more constraints.

The only nonvanishing Poisson brackets among the constraints are
$$\{\phi_i^1(x),\phi_j^1(y)\}=-\frac{\alpha}{2\pi}\varepsilon_{ij}\delta(x-y),
\quad \{\phi_i^1(x),\phi_3^2(y)\}=\frac{\alpha}{2\pi}\varepsilon_{mi}
\partial_m\delta(x-y).$$
Therefore, the matrix ~${\bf W}=\left\|\{\phi_\alpha^{m_\alpha},\phi_\beta^
{m_\beta}\}\right\|$ takes the form:
\be \label{Chern:W}
{\bf W}=\frac{\alpha}{2\pi}\left(\begin{array}{rrrr}
0~ & -1~ & ~~0 & -\partial_2 \\
1~ & 0~ & ~~0 & \partial_1 \\
0~ & 0~ & ~~0 & 0~ \\
\partial_2 & -\partial_1 & ~0  & 0~
\end{array}  \right)\delta(x-y).
\ee
>From ~$\mbox{rank}{\bf W}=2$~ we conclude that two constraints are of second
class and the two ones are of first class.

With the help of the transformation
$$\bar \phi_3^2 = \phi_3^2+c_1\phi_1^1+c_2\phi_2^1$$
we shall satisfy the equality ~~$\{\bar \phi_3^2,\phi_i^1\}=0$~~ if ~~$c_i=-
\partial_i$.

Thus, we obtain the canonical set of constraints: $\Psi_1^1=\phi_1^1,$
$\Psi_2^1=\phi_2^1,~ \Phi_1^1=\phi_3^1,~ \Phi_1^2=\bar \phi_3^2=j_\varphi,$
separated into the ones of first and second class, since now the matrix ~${\bf
W}$~ has the form:  $$ {\bf W}^\p=\frac{\alpha}{2\pi}\left(\begin{array}{rrrr}
0 & -1 & ~~0 & ~~0 \\
1 & 0 & ~~0 & ~~0 \\
0 & 0 & ~~0 & ~~0 \\
0 & 0 & ~~0 & ~~0
\end{array} \right)\delta(x-y).$$
Further, we seek the generator $G$ in the form
\be \label{Chern:G-Dirac}
G=\int d^2x \Bigl[\eta_1^1~\Psi_1^1+\eta_2^1~\Psi_2^1+\varepsilon_1^1~\Phi_1^1+
\varepsilon_1^2~\Phi_1^2\Bigr].
\ee
>From the second condition \rf{PB-G-Psi1} of conservation of $\Sigma_1$ under
transformations \rf{q-q^prime} we derive ~$\eta_1^1=\eta_2^1=0,$  i.e. the
constraints of second class do not contribute to $G$. The first condition
\rf{PB-G-Phi1} of conservation of $\Sigma_1$ is realized because
$$\bigl\{\Phi_1^1,\Phi_1^2\bigr\}=0.$$
Since $g_{1~1}^{1~2}=1$ and $g_{1~1}^{2~2}=0$ in
\rf{PB-Phi-H^prime}, eq.\rf{eq:eps} accepts the form:
$$\dot \varepsilon_1^2 +\varepsilon_1^1 = 0,$$
i.e. ~$\varepsilon_1^1=-\dot \epsilon(x)$~ where $\epsilon(x)\equiv
\varepsilon_1^2$.~ Therefore we obtain
\be \label{Chern:G}
G=\int d^2x \Bigl\{-\dot \epsilon\pi_0 + \epsilon\bigl[~i(\varphi\pi_\varphi
-\varphi^* \pi_{\varphi^*})-\partial_i\pi_i \bigr]\Bigr\},
\ee
from which it is easily to derive the local-symmetry transformations in the
phase space:
\bea \label{Chern:qp}
\left.\begin{array}{llll}
\delta\varphi(x)=i\epsilon(x)\varphi(x), &~~
\delta\pi_\varphi(x)=-i\epsilon(x)\pi_\varphi(x),           \\
\delta\varphi^*(x)=-i\epsilon(x)\varphi^*(x), &~~
\delta\pi_{\varphi^*}(x)=i\epsilon(x)\pi_{\varphi^*}(x), \\ \delta
A_0(x)=\dot\epsilon(x), &~~ \delta\pi_0(x)=0, \\ \delta
A_i(x)=\partial_i\epsilon(x), &~~
\delta\pi_i(x)=0.
\end{array}  \right.
\eea
With the help of \rf{q-dot-q} it is easily to write the local-symmetry
transformations in the $(q,\dot q)$-space and to obtain that
~$\delta{\cal L}=\partial_\mu\bigl[\frac{\alpha}{4\pi}\varepsilon^
{\mu\nu\lambda}\epsilon(x)\partial_\nu A_\lambda\bigr],$~ i.e. the theory is
quasi-invariant under obtained transformations.

4. Now we consider the well-known case of spinor electrodynamics:
\be \label{el.dyn:L}
{\cal L}=-\frac{1}{4} F_{\mu\nu} F^{\mu\nu}+i\overline{\psi}\gamma_\mu(\partial_\mu
-ieA_\mu)\psi-m\overline{\psi} \psi
\ee
where ~~$ F_{\mu\nu}=\partial_\mu A_\nu-\partial_\nu A_\mu.$~ Here~ $A_\mu,
\psi,\overline{\psi}$~ play the role of the generalized coordinates. The
generalized momenta are
$$\pi_\mu=\frac{\partial L}{\partial \dot A_\mu}=F_{0\mu}, \quad
p_\psi=\frac{\partial L}{\partial \dot \psi}=i\overline{\psi}\gamma_0, \quad
p_{\overline{\psi}}=\frac{\partial L}{\partial \dot {\overline{\psi}}} =0,$$
from which we have three primary constraints:
\be \label{el.dyn:phi-prim}
\phi_1^1=\pi_0, \qquad \phi_2^1=p_\psi-i\overline{\psi}\gamma_0, \qquad
\phi_3^1=p_{\overline{\psi}}
\ee
and the total Hamiltonian:
\bea \label{el.dyn:H_T}
H_T &=& \int d^3x \Bigl[~\frac{1}{4} F_{ij} F^{ij}+\frac{1}{2}\pi^i
\pi^i+\pi_i\partial_i A_0+ iep_\psi A_0 \psi  \nn\\ & &
~~~+i\overline{\psi}\gamma_i(\partial_i-ieA_i)\psi+m\overline{\psi} \psi
+u_1\phi_1^1+u_2\phi_2^1+u_3\phi_3^1~\Bigr].
\eea
Among the conditions of the constraint conservation in time~ $\dot{\psi}_i^1=
0~(i=1,2,3)$~ the two last ones serve for determining the Lagrangian
multipliers $u_2$ and $u_3$. From the first condition we obtain one secondary
constraint $$\phi_1^2= \partial_i\pi^i-iep_\psi \psi,$$ and there do not arise
more constraints. Calculating the matrix ~${\bf W}=\left\|
\{\phi_\alpha^{m_\alpha},\phi_\beta^{m_\beta}\}\right\|$:
\be \label{el.dyn:W}
{\bf W}=\delta(x-x^\p)\left(\begin{array}{cccc}
0 & ~~0 & 0 & 0 \\
0 & ~~0 & i\gamma_0 & -iep_\psi \\
0 & ~~i\gamma_0 & 0 & -e\gamma_0 \psi \\
0 & ~~iep_\psi & e\gamma_0 \psi & 0
\end{array}  \right),
\ee
we see that ~$\mbox{rank}{\bf W}=2$; therefore, two constraints are of second
class and the two ones are of first class. Now implementing our procedure, we
shall pass to the canonical set of constraints by the equivalence
transformation:
\begin{displaymath}
\left(\begin{array}{c}
\Psi_1^1\\ \Psi_2^1\\ \Phi_1^1\\ \Phi_1^2 \end{array}  \right)=
\left(\begin{array}{cccc}
1 & 0 & ~0 & ~~~0 \\
0 & 1 & ~0 & ~~~0 \\
0 & 0 & ~1 & ~~~0 \\
ie\psi & -ie\overline{\psi} & ~0 & ~~~1
\end{array}  \right)
\left(\begin{array}{c}
\phi_2^1\\ \phi_3^1\\ \phi_1^1\\ \phi_1^2 \end{array}  \right)=
\left(\begin{array}{c}
p_\psi-i\overline{\psi}\gamma_0\\ p_{\overline{\psi}}\\ \pi_0\\
\partial_i\pi^i-ie(p_\psi \psi+\overline{\psi}p_{\overline{\psi}})
\end{array}  \right)
\end{displaymath}
where the constraints are already separated into the ones of first and second
class, since now the matrix ~${\bf W}$~ has the form:
$$ {\bf W}^\p=\delta(x-x^\p)\left(\begin{array}{cccc}
0 & i\gamma_0 & ~0 & ~~0 \\
i\gamma_0 & 0 & ~0 & ~~0 \\
0 & 0 & ~0 & ~~0 \\
0 & 0 & ~0 & ~~0
\end{array} \right).$$
Further, we look for the generator $G$ in the form
\be \label{el.dyn:G-Dirac}
G=\int d^3x \Bigl[\eta_1^1~\Psi_1^1+\eta_2^1~\Psi_2^1+\varepsilon_1^1~\Phi_1^1+
\varepsilon_1^2~\Phi_1^2\Bigr].
\ee
>From the second condition \rf{PB-G-Psi1} of conservation of the
primary-constraints surface $\Sigma_1$ under transformations \rf{q-q^prime} we
derive ~$\eta_1^1=\eta_2^1=0,$  i.e. the constraints of second class do not
contribute to $G$. The first condition \rf{PB-G-Phi1} of conservation of
$\Sigma_1$ is realized because $$\bigl\{\Phi_1^1,\Phi_1^2\bigr\}=0.$$
Taking into account that $g_{1~1}^{1~2}=-1$ and $g_{1~1}^{2~2}=0$ in
\rf{PB-Phi-H^prime}, eq.\rf{eq:eps} accepts the form:
$$\dot \varepsilon_1^2 -\varepsilon_1^1 = 0,$$
i.e. ~$\varepsilon_1^1=\dot \varepsilon$~ where $\varepsilon\equiv
\varepsilon_1^2$.~ Therefore we have
$$G=\int d^3x \Bigl\{\dot \varepsilon \pi_0+\varepsilon\bigl[\partial_i \pi^i-
ie(p_\psi \psi+ \overline{\psi}p_{\overline{\psi}})\bigr]\Bigl\},$$ from which
it is easily to obtain the gauge transformations in the phase space and
well-known transformation rule:  $$\delta A_\mu=\partial_\mu \varepsilon,\quad
\delta \psi=ie\varepsilon\psi, \quad \delta
\overline{\psi}=-ie\varepsilon\overline{\psi}.$$ 

\section{Conclusion}

Constrained special-form theories with first- and second-class constraints,
when the first-class primary constraints are the ideal of quasi-algebra of all
the first-class constraints, are considered. One must say that this
restriction on the algebra of constraints is fulfilled in most of the
physically interesting theories, e.g., in electrodynamics, in the Yang --
Mills theories, etc., and it has been used by us in previous works \cite{GSST}
in the case of dynamical systems only with the first-class constraints and
also by other authors in obtaining gauge transformations on the basis of
different approaches \cite{Castellani,Cawley,Cabo,Bergmann,Zanelli}.

Here in the framework of generalized Hamiltonian formalism by Dirac for systems
with first- and second-class constraints we have suggested the method of
constructing the generator of local-symmetry transformations in both phase and
configuration space. The generator is derived from the requirement of
quasi-invariance (within a surface term) of the action functional (in the
phase space) under desired transformations which must be supplemented by the
demand on the primary-constraints surface $\Sigma_1$ to be conserved at these
transformations. Necessity of second requirement can be seen from following
reasoning. Because $\Sigma_1$ is whole $(q,\dot q)$-space image in the phase
space and under operation of the local-symmetry transformation group the
$(q,\dot q)$-space is being mapped into itself in a one-to-one manner, then
one-to-one mapping of $\Sigma_1$ into itself corresponds to this in the phase
space.

Note that the condition of the $\Sigma_1$ conservation actually is not the
additional restriction on the properties of the local-symmetry transformation
generator. It naturally follows from definition of the symmetry group of
the action functional (see the explanation after relation \rf{PB-G-Phi1}).

It is shown that second-class constraints do not contribute to the
local-symmetry transformation law and do not generate global transformations
in lack of first-class constraints.

The corresponding transformations of local symmetry in the $(q,\dot q)$-space
are determined with the help of formulae \rf{q-dot-q}.

When deriving the local-symmetry transformation generator the employment of
obtained equation system \rf{eq:eps} is important, the solution of which
manifests a mechanism of appearance of higher derivatives of coordinates and
group parameters in the Noether transformation law in the configuration spaÓe,
the highest possible order of coordinate derivatives being determined by the
structure of the first-class constraint algebra, and the order of the highest
derivative of group parameters in the transformation law being by unity
smaller than the number of stages in deriving secondary constraints of first
class by the Dirac procedure. The arising problem of canonicity of
transformations in the phase space in the presence of higher derivatives of
coordinates and momenta will be considered in our subsequent paper.

So, we can state in the case of special-form theories with first- and
second-class constraints that the necessary and sufficient condition for
certain quantity $G$ to be the local-symmetry transformation generator is the
representation of $G$ as the linear combination of all the first-class
constraints (and only of them) with the coefficients determined by the system
of equations \rf{eq:eps}.

Obtained generator \rf{G} satisfies the group property \rf{G:group}. The
amount of group parameters, which determine a rank of quasigroup of these
transformations, equals to the number of primary constraints of first class.

As it is known, gauge-invariant theories belong to the class of degenerate
theories. In this paper we have shown that the degeneracy of special-form
theories with the first- and second-class constraints is due to their
quasi-invariance under local-symmetry transformations.

\section*{Acknowledgments}
One of the authors (S.A.G.) thanks the Russian Foundation for Fundamental
Research (Grant N$^{\underline {\circ}}$ 96-01-01223) for support.

\end{document}